\documentclass[graybox]{svmult}

\usepackage{mathptmx}       
\usepackage{helvet}         
\usepackage{courier}        
\usepackage{type1cm}        
\usepackage{makeidx}         
\usepackage{graphicx}        
\usepackage{multicol}        
\usepackage[bottom]{footmisc}

\makeindex             

\newcommand{\methane}{CH$_4$}
\newcommand{\nitrogen}{N$_2$}
\newcommand{\ammonia}{NH$_3$}
\newcommand{\methyl}{CH$_3$}
\newcommand{\hydrogen}{H$_2$}
\newcommand{\ethane}{C$_2$H$_6$}
\newcommand{\diacet}{C$_4$H$_2$}

\newcommand{\ethylene}{C$_2$H$_4$}
\newcommand{\cyanoacet}{HC$_3$N}
\newcommand{\cyanogen}{C$_2$N$_2$}
\newcommand{\methylacet}{C$_3$H$_4$}
\newcommand{\propane}{C$_3$H$_8$}
\newcommand{\coo}{CO$_2$}
\newcommand{\benzene}{C$_6$H$_6$}
\newcommand{\acetonitrile}{CH$_3$CN}
\newcommand{\water}{H$_2$O}
\newcommand{\nratio}{$^{14}$N/$^{15}$N}

\newcommand{\cm}{cm$^{-1}$}
\newcommand{\micron}{$\mu$m}
\newcommand{\hcnfifteen}{HC$^{15}$N}
\newcommand{\dg}{$^{\circ}$}

\begin{document}

\title*{Nitrogen in the stratosphere of Titan from Cassini CIRS infrared spectroscopy}
\author{Conor A. Nixon, Nicholas A. Teanby, Carrie M. Anderson and Sandrine Vinatier}
\institute{Conor A. Nixon \at Department of Astronomy, University of Maryland, College Park, MD 20742, USA, \email{conor.a.nixon@nasa.gov}
\and Nicholas A. Teanby \at School of Earth Sciences, University of Bristol, Wills Memorial Building, Queen's Road, Bristol, BS8 1RJ, UK, \email{n.teanby@bristol.ac.uk}
\and Carrie M. Anderson \at Planetary Systems Laboratory, NASA Goddard Space Flight Center, Greenbelt, MD 20771, USA \email{carrie.m.anderson@nasa.gov}
\and Sandrine Vinatier \at LESIA, Observatoire de Paris, CNRS, 5 Place Jules Janssen, 92195 Meudon Cedex, France \email{sandrine.vinatier@obspm.fr}}
%
%
\maketitle

\abstract*{In this chapter we describe the remote sensing measurement of nitrogen-bearing species in Titan's atmosphere by the Composite Infrared Spectrometer (CIRS) on the Cassini spacecraft. This instrument, which detects the thermal infrared spectrum from 10--1500~\cm\ (1000--7~$\mu$m) is sensitive to vibrational emissions of gases and condensates in Titan's stratosphere and lower mesosphere, permitting the measurement of ambient temperature and the abundances of gases and particulates. Three N-bearing species are firmly detected: HCN, HC$_3$N and C$_2$N$_2$, and their vertical and latitudinal distributions have been mapped. In addition, ices of HC$_3$N and possibly C$_4$N$_2$ are also seen in the far-infrared spectrum at high latitudes during the northern winter. The HC$^{15}$N isotopologue has been measured, permitting the inference of the \nratio\ in this species, which differs markedly (lower) than in the bulk nitrogen reservoir (\nitrogen ). We also describe the search in the CIRS spectrum, and inferred upper limits, for \ammonia\ and \acetonitrile . CIRS is now observing seasonal transition on Titan and the gas abundance distributions are changing accordingly, acting as tracers of the changing atmospheric circulation. The prospects for further CIRS science in the remaining five years of the Cassini mission are discussed.}

\abstract{In this chapter we describe the remote sensing measurement of nitrogen-bearing species in Titan's atmosphere by the Composite Infrared Spectrometer (CIRS) on the Cassini spacecraft. This instrument, which detects the thermal infrared spectrum from 10--1500~\cm\ (1000--7~$\mu$m) is sensitive to vibrational emissions of gases and condensates in Titan's stratosphere and lower mesosphere, permitting the measurement of ambient temperature and the abundances of gases and particulates. Three N-bearing species are firmly detected: HCN, HC$_3$N and C$_2$N$_2$, and their vertical and latitudinal distributions have been mapped. In addition, ices of HC$_3$N and possibly C$_4$N$_2$ are also seen in the far-infrared spectrum at high latitudes during the northern winter. The HC$^{15}$N isotopologue has been measured, permitting the inference of the \nratio\ in this species, which differs markedly (lower) than in the bulk nitrogen reservoir (\nitrogen ). We also describe the search in the CIRS spectrum, and inferred upper limits, for \ammonia\ and \acetonitrile . CIRS is now observing seasonal transition on Titan and the gas abundance distributions are changing accordingly, acting as tracers of the changing atmospheric circulation. The prospects for further CIRS science in the remaining five years of the Cassini mission are discussed.}

\section{Introduction}
\label{sect:intro}

The two principal gaseous components of Titan's atmosphere are molecular nitrogen \nitrogen\ (98.5\% in the stratosphere) and methane \methane\ (1.4\% in stratosphere) \cite{niemann10}. From just these two molecules and three elements a wealth of organic chemistry develops,\footnote{There are also some oxygen species, most notably CO at 50 ppm, which are not discussed here.} resulting in a plethora of hydrocarbons and nitriles. The true chemical complexity of Titan's atmosphere was first revealed in 1980 with the Voyager encounter at Titan, during which a number of new molecular species were first detected through infrared spectroscopy by the IRIS (Infrared Interferometer Spectrometer) instrument: HCN, \ethylene\ and \ethane\ \cite{hanel81}; \cyanoacet , \cyanogen\ and \diacet\ \cite{kunde81}; and the C$_3$ molecules \methylacet\ and \propane\ \cite{maguire81}. Further stratospheric species were found later: \coo , CO, \acetonitrile , \benzene\ and \water\ \cite{nixon10b}. 

More recently, the NASA Cassini spacecraft entered Saturn orbit in July 2004, and has since made more than 80 close flybys of Titan at the time of writing. Equipped with an Ion and Neutral Mass Spectrometer (INMS) \cite{waite04}, Cassini has directly sampled the upper atmosphere of Titan, finding further chemical diversity, including ammonia (\ammonia ), amines (-NH$_2$) and imines (-NH), and heavier hydrocarbons such as toluene (C$_7$H$_8$) \cite{vuitton07}. In the stratosphere, Cassini's infrared spectrometer CIRS (Composite Infrared Spectrometer) \cite{flasar04b} has mapped the vertical, lateral and temporal variations of the stratospheric nitriles, which constitutes Sect. \ref{sect:nitriles} of this chapter. Later sections cover related topics: nitrile condensates and ices in Sect. \ref{sect:ices}, the isotopic ratio \nratio\ as measured in HCN in Sect. \ref{sect:isotopes}, and the search for molecules as yet undetected in the stratosphere in Sect. \ref{sect:search}.

We begin this chapter by briefly reviewing chemical processes from the break-up of \nitrogen\ to the formation of H$_x$C$_y$N$_z$ species, and also the CIRS instrument that is the source of the scientific information presented here.

\subsection{Nitrogen chemistry}
\label{sect:chem}

In Titan's upper atmosphere, nitrogen chemistry begins when molecular nitrogen is dissociated/ionized by solar UV radiation\footnote{The chemical discussion in this section follows the more detailed description in \cite{lavvas08a}}:

\begin{eqnarray}
{\rm N_2} + h\nu & \rightarrow & {\rm N(^4S) + N^+ + e^- \hspace*{11mm} \lambda < 510 \AA \hspace*{5mm} (10\%) }\\
 \: & \rightarrow & {\rm N(^2D) + N^+ + e^- \hspace*{10mm} \lambda < 510 \AA \hspace*{5mm} (90\%) } \\
 \: & \rightarrow & {\rm N_2^+ + e^- \hspace*{23mm} 510 < \lambda < 796 \AA}\\
 \: & \rightarrow & {\rm N(^2D) + N(^4S) \hspace*{13mm} 796 < \lambda < 1000 \AA} 
\end{eqnarray}

In addition, high-energy Galactic Cosmic Rays (GCRs, mainly protons and alpha particles) reach the lower atmosphere, where a second dissociation region occurs peaking near 100~km. To form nitriles or amines/imines, carbon and hydrogen are required, which are supplied by the solar photodissociation of methane:

\begin{eqnarray}
{\rm CH_4} + h\nu & \rightarrow & {\rm CH + H + H_2 \hspace*{12mm} \lambda \leq 145 nm } \hspace*{5mm} (7\%) \\
 & \rightarrow & {\rm  ^1CH_2 +  H_2  \hspace*{15mm} \lambda \leq 145 nm } 
 \hspace*{5mm} (58.4\%) \\
 & \rightarrow & {\rm ^1CH_2 + 2H  \hspace*{15mm} \lambda \leq 145 nm } 
 \hspace*{5mm} (5.5\%) \\
 & \rightarrow & {\rm CH_3 + H  \hspace*{18mm} \lambda \leq 145 nm } 
 \hspace*{5mm} (29.1\%) \\
 & \rightarrow & {\rm CH_3 + H   \hspace*{18mm} \lambda > 145 nm } 
 \end{eqnarray}

Nitriles arise when \methyl\ reacts with N, either by:

\begin{equation}
{\rm N + CH_3 \rightarrow HCN + H_2 }
\label{eq:hcn}
\end{equation}

\noindent or

\begin{eqnarray}
{\rm N + CH_3} & \rightarrow & {\rm H_2CN + H} \\
{\rm H_2CN + H} & \rightarrow & {\rm HCN + H_2}
\end{eqnarray}

\noindent
which has the same net effect as reaction (\ref{eq:hcn}) but an overall faster rate. HCN may be photolyzed to H + CN, and the CN radical will then either (a) catalytically destroy unsaturated hydrocarbons and \hydrogen\ by H-abstraction, or (b) form heavier nitrile species by substitution of CN for H, e.g. \ethylene\ becomes C$_2$H$_3$CN (acrylonitrile or vinyl cyanide).

The production of amines and imines begins when dissociated nitrogen atoms interact with methane:

\begin{eqnarray}
{\rm N(^2D) + CH_4} &  \rightarrow & {\rm NH + CH_3 \hspace*{1.3cm} (20\%) } \\
 & \rightarrow & {\rm CH_2NH + H \hspace*{1cm} (80\%) }
 \end{eqnarray}

The self-reaction of two imidogen radicals then leads to the amino radical, or alternatively atomic nitrogen plus molecular hydrogen leads to the same result:

\begin{eqnarray}
{\rm 2 NH} & \rightarrow & {\rm NH_2 + N} \\
{\rm N + H_2 + M} & \rightarrow & {\rm NH_2 + M}
\end{eqnarray}

Finally, ammonia is produced from the amino radical plus atomic hydrogen:

\begin{equation}
{\rm NH_2 + H} \rightarrow {\rm NH_3}
\end{equation}

Other pathways are possible to these products: we have summarized only the (believed to be) principal pathways here. For a fuller description see \cite{lavvas08a}. In the following sections of this chapter, we will describe how the concentrations of various nitrile species have been measured by CIRS in the middle atmosphere, and on the search for ammonia in the stratosphere. 

\subsection{The Composite Infrared Spectrometer (CIRS)}
\label{sect:CIRS}

Cassini CIRS is a dual interferometer, comprised of separate far-infrared and mid-infrared spectrometers sharing a common telescope, foreoptics, reference laser, scan mechanism and other sub-systems to save mass and size. The telescope is a Cassegrain type, with 508 mm beryllium primary and 76 mm secondary mirrors. The incident beam is field-split and sent to either the far-IR or mid-IR interferometer. The far-IR interferometer is a Martin-Puplett (polarizing) type, using wire-grid polarizers to amplitude split radiation between 10--600~\cm\ (1000--17~\micron ) in wavenumber (wavelength). The time-varying interferogram signal produced by the scanning of one retroreflector is afterwards detected by a thermopile (bolometer) detector known as FP1 (Focal Plane 1), which is 1 mm in size and has apparent field-of-view projected on the sky plane of 2.5 mrad FWHM (full-width to half-maximum of Gaussian response). 

The mid-IR interferometer is a standard Michelson type covering the spectral range 600-1400~\cm\ (17--7~\micron ), and the signal is detected by one of two arrays. CIRS FP3 (Focal Plane 3) consists of a $1\times 10$ array of photoconductive (PC) detectors sensitive from 600-1100~\cm\ (17--9~\micron ), while FP4 (Focal Plane 4) is a similar $1\times 10$ array of photovoltaic detectors sensitive from 1100--1400~\cm\ (9--7~\micron ). Pixels of both arrays have square-shaped projected fields-of-view 0.273 mrad in width. By varying the travel distance (or scan time) of the mirror carriage, spectral resolutions varying from 0.5--15.5~\cm\ are possible, after application of Hamming apodization to reduce the ripples/ringing caused by the finite Fourier Transform.

The FP1 detector is held at 170~K, identical to the rest of the instrument optics and mechanical assembly, while FP3/FP4 are cooled to an operating temperature of $\sim$75~K via a 30-cm radiator pointed at cold space. For calibration therefore, the FP1 detector needs only one temperature reference target (space at 2.73~K), while FP3/FP4 require both reference scans of deep space and also an internal warm shutter at $\sim$170~K. A more detailed overview of the instrument can be found in the literature \cite{flasar04b, kunde96}, while more detailed descriptions of the FP1 far-IR interferometer and the FP3/FP4 mid-IR interferometer have been separately published \cite{brasunas04,nixon09a}.

As Cassini approaches Titan, CIRS performs different science observations depending on distance, based on the differing capabilities of the mid-IR and far-IR spectrometers. At large distances (380,000--260,000~km, 19--13 hrs from closest approach), the mid-IR arrays are typically used in nadir-scanning mode, where the arrays are `combed' across Titan's visible disk. Closer in (260,000-180,000~km, 13--9 hrs), the FP1 detector is placed on the disk and long-integrations are made at the fixed location, to build up high S/N at the highest spectral resolution (0.5~\cm ) to search for/measure weak species in the far-IR. At medium distances (180,000--100,000~km, 9--5~hrs), the mid-IR detectors acquire sufficient spatial resolution to resolve the limb of Titan's atmosphere (defined by the atmospheric scale height, about 50~km), and are used for limb mapping or integrating. Less than 5 hours (100,00~km) from Titan, the far-IR channel again takes precedence, used first in nadir mapping mode (100,000--45,000~km, 5:00--2:15~hrs) and finally in limb mode (45,000--5,000~km, 2:15--0:15~hrs), when the spatial resolution becomes sufficient to resolve the limb using the large FP1 detector. Further details of the Titan science strategy can be found in existing publications \cite{flasar04b, nixon10c, nixon12a}.

\section{Nitrile species in the stratosphere}
\label{sect:nitriles}

\begin{figure}[b]
\includegraphics[scale=.48]{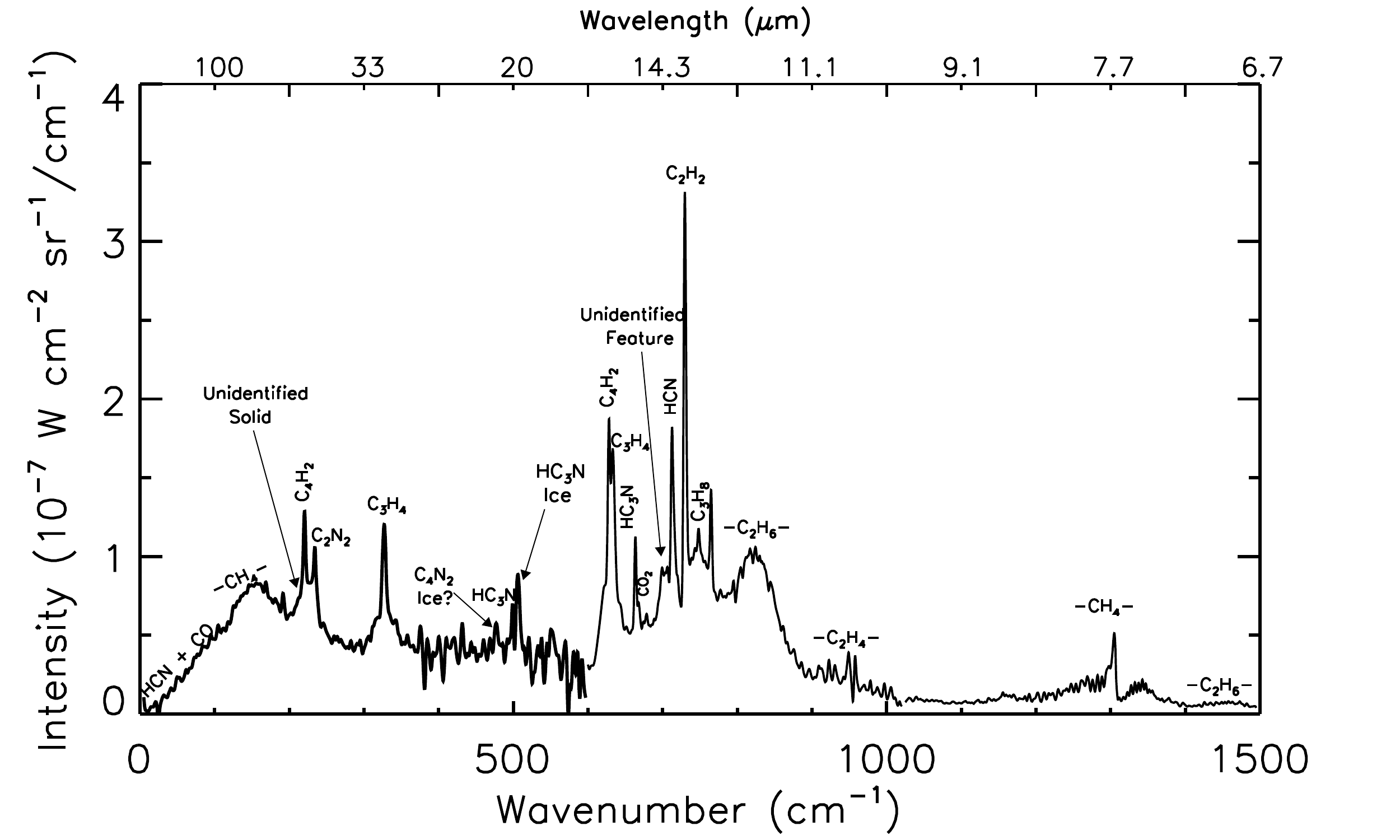}
\caption{Sample CIRS limb spectrum of Titan at 70\dg N, 125-160~km altitude from August 2007. Vibrational bands of HCN, HC$_3$N and C$_2$N$_2$ are labeled, along with an HC$_3$N ice signature. Adapted from \cite{Anderson11}.}
\label{fig:spec}
\end{figure}

The three major nitrile species in Titan's atmosphere are HCN, HC$_3$N, and C$_2$N$_2$, and all are spectroscopically detected and measured by CIRS  - see Fig. \ref{fig:spec}. Through modeling these emissions, and assuming the atmospheric temperature is independently known, the abundances can be measured. In this section we discuss the inferred global distribution of these nitriles and consider what this tells us about nitrogen-species chemistry on Titan. 

\subsection{Vertical profiles at low latitudes}
\label{sect:vertical}

Nitrile species are produced by photochemical reactions in Titan's upper atmosphere at altitudes above 500~km (Sect.~\ref{sect:chem}). This is above the region that is observable with Cassini CIRS. However, these species are transported into the lower mesosphere and stratosphere by vertical mixing processes, which allow them to be observed. At altitudes of approximately 100--150~km in the lower stratosphere, temperatures become cold enough to allow condensation and rain-out of nitrile species. This source-sink arrangement sets up positive vertical concentration gradients where the relative abundance increases with altitude.Photochemical models predict that vertical gradients are steeper for species with shorter photochemical lifetimes (e.g. HC$_3$N) as they have less time to mix into lower atmospheric levels before being destroyed by photolysis.

\begin{figure}[t]
\sidecaption
\includegraphics[scale=.5]{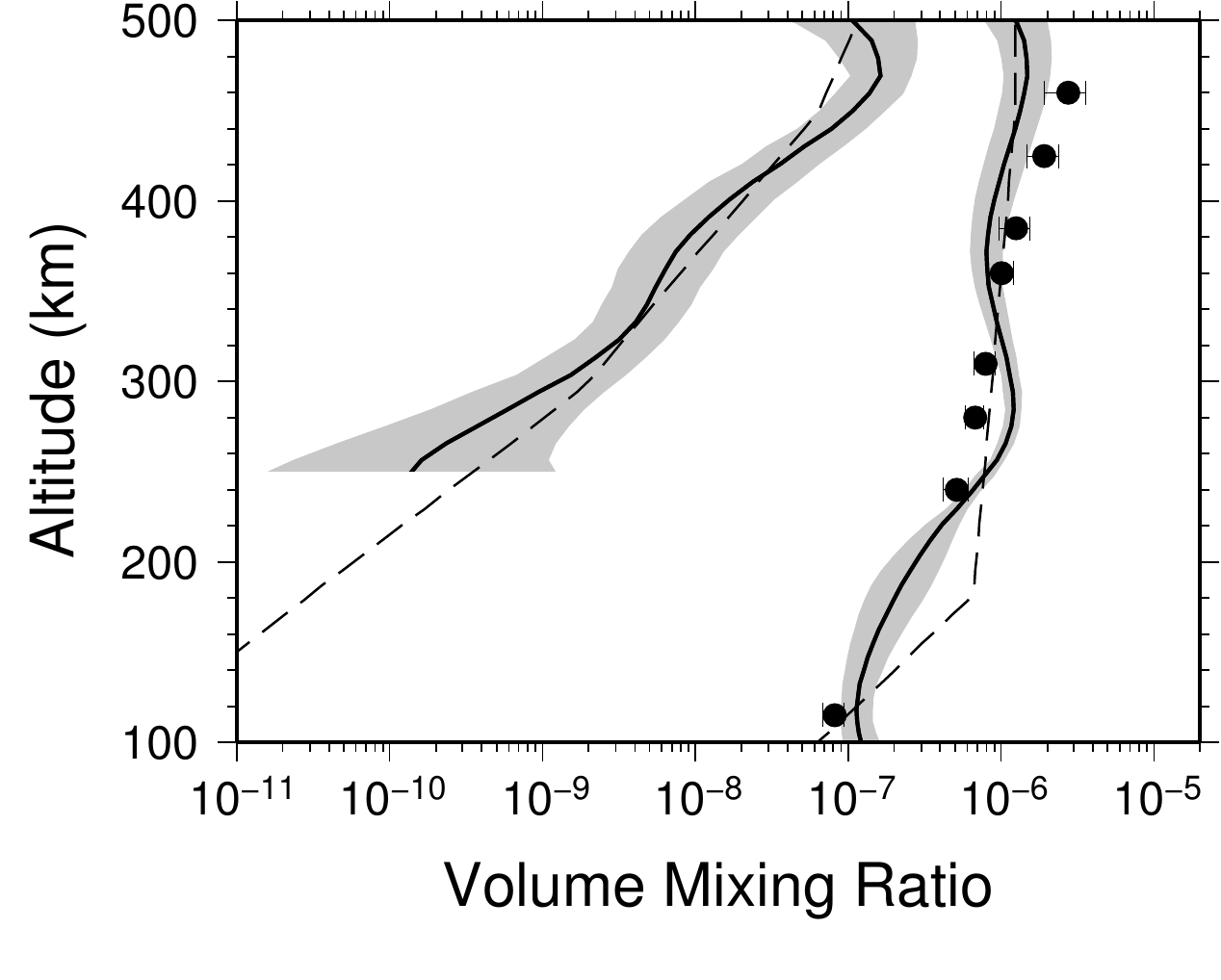}
\caption{Profiles of HCN and HC$_3$N from \cite{07teaetal} (solid line with grey error envelope) and \cite{07vinetal} (filled circles with error bars) derived from CIRS limb observations in December 2004 (Tb flyby) at a latitude of 15$^{\circ}$S. Dashed lines show the previous estimates from ground-based work \cite{02maretal}. Agreement between studies is good and shows a steep gradient for HCN and a very steep gradient for HC$_3$N.}
\label{fig:profiles} 
\end{figure}

Shortly after Cassini's arrival at the Saturnian system in July 2004, CIRS limb sounding observations from the early flybys gave the first detailed vertical profiles of nitriles in Titan's atmosphere \cite{07teaetal,07vinetal} (Fig.~\ref{fig:profiles}). The CIRS results confirm earlier ground based sub-millimeter work \cite{97hidetal,02maretal,04gur} based on high-resolution measurements of the lineshape of nitrile rotational lines. Both HCN and HC$_3$N were observed to have steep vertical gradients - in broad agreement with photochemical models. Titan's equatorial abundances had the best agreement with the Earth-based results because of the sampling bias towards low latitudes in disk-averaged ground-based spectra. Equatorial latitudes also bear the most resemblance to photochemical model predictions, as the equator is closest to equilibrium conditions in terms of minimal atmospheric motion and ambient solar flux. The abundance of \cyanogen\ at the equator was measured to be 0.055 ppb at 125~km \cite{09teaetal_c2n2}. Comparisons with photochemical models \cite{04wilatr} suggests an abundance this high is most consistent with \nitrogen\ dissociation by cosmic rays in the lower stratosphere.

Interestingly, early results showed that the gradient for HCN was steeper than predicted by current photochemical models. This suggests an additional sink for nitriles in lower atmosphere, effectively reducing their atmospheric lifetimes compared to photochemical model predictions.

\subsection{Global variations of abundances}
\label{sect:global}

The early CIRS results suggest that Titan's nitriles are more complex than previously thought.
Further insight into the chemical processes can be obtained by considering the distribution of nitriles across Titan's globe - a feat now possible because of the spatial sampling ability of CIRS.

At the start of the mission in 2004 Titan was experiencing early northern winter. The resulting differential heating between southern and northern hemispheres gives rise to large convection cells - analogous to the Hadley cells on Earth - which causes upwelling in the south and subsidence in the north. These vertical atmospheric motions have the ability to modify the equilibrium photochemical profiles \cite{01lebetal,09teaetal_rs}. Where subsidence is occurring, the entire vertical profile is advected downwards. As a consequence of the positive vertical gradient, this leads to an increased abundances at each altitude. Conversely, upwelling causes a decrease in abundance at each altitude. This phenomenon helps us in two ways:

\begin{enumerate}
\item By observing the increase and decrease in abundances across Titan's globe, we are able to map out vertical circulation patterns.
\item By looking at variations in subsidence-induced enrichment between different gases we can learn about the relative vertical gradients - as steeper vertical gradients will result in greater enrichment. This in turn informs us about the relative lifetimes of the different species.
\end{enumerate}

\begin{figure}[t]
\sidecaption
\includegraphics[scale=.3]{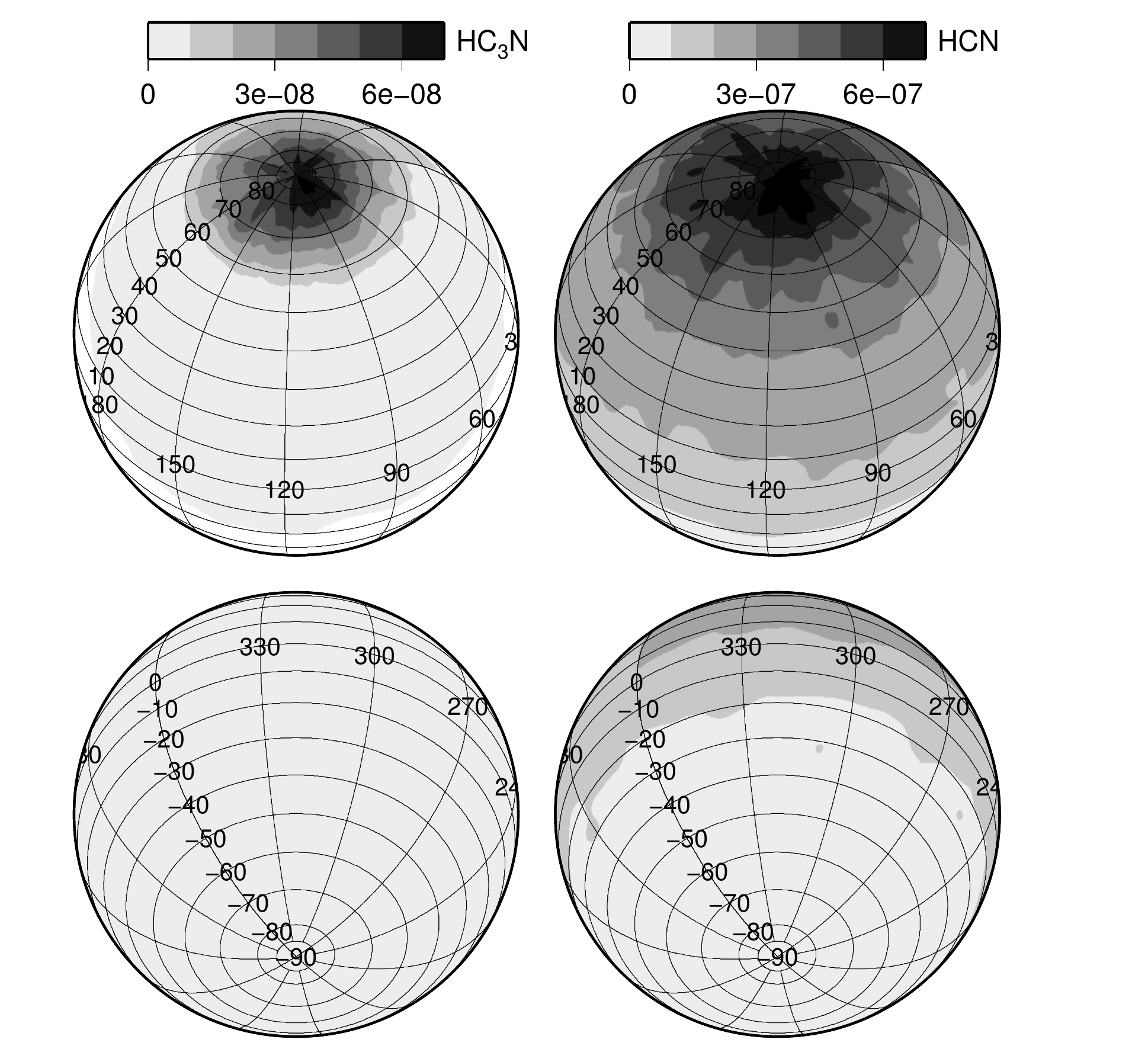}
\caption{Maps of HCN and HC$_3$N derived for the northern and southern hemispheres. HC$_3$N is greatly enriched at high northern latitudes, but decreases sharply south of 60$^{\circ}$N and is present at very low levels elsewhere. HCN is also enriched in the north, but has a more gradual variation from south to north. These observations suggest subsidence at the north pole and are consistent with the presence of a polar vortex. Redrawn from \cite{08teaetal}.}
\label{fig:maps} 
\end{figure}

The best way to map the global distribution of the different gases is by using nadir (downward looking) observation sequences. CIRS medium spectral resolution mapping scans are particularly useful for this as they observe an entire hemisphere at once and are taken on nearly every flyby.
High spectral-resolution north-south swaths, although less numerous, are also extremely useful as they provide better signal-to-noise and spectral discrimination between different gases.
Nadir datasets were used early in the mission \cite{05flaetal,06teaetal,07couetal} to map out the global distribution of many of Titan's trace species - including the nitriles.
The early studies have subsequently been expanded \cite{09teaetal_rs,08teaetal,08teaetal_jgr,09teaetal_c2n2,10couetal,10teaetal_faraday,10teaetal_apj,10vinetal} and the northern winter distribution of most trace species is now well understood.
Fig~\ref{fig:maps} shows the distribution of HCN and HC$_3$N from a typical northern winter flyby.

A large enrichment of most trace species - including nitriles - is observed at the north pole compared to the rest of the planet - indicating a large single south-to-north circulation cell with subsidence in the north, consistent with the winter season.
Detailed studies of the north pole have shown there to be a strong circumpolar vortex \cite{05flaetal,08teaetal_jgr,08achetal}.
The vortex is produced by a combination of poleward transport of air by the atmospheric circulation cells and conservation of angular momentum, which causes a spin up of stratospheric and mesospheric air masses.
Vortex wind speeds were determined using the thermal wind equation, which implied wind speeds of nearly 200~m/s in the stratosphere at around 30--60$^{\circ}$N.
Across the vortex boundary at 60$^{\circ}$N, there is a strong potential vorticity gradient, which indicates that a significant mixing barrier exists separating polar and non-polar air masses.
Enhanced infrared emission from large abundances of trace gases such as HCN then act as an effective cooling mechanism, which further reinforces the general circulation. The chemistry and dynamics in this region are thus intimately linked.

Fig.~\ref{fig:maps} shows that north-polar HC$_3$N enrichment is much greater than that of HCN.
This is consistent with the much shorter photochemical lifetime of HC$_3$N (0.8~years) compared to HCN (44~years) \cite{04wilatr}. HC$_3$N is also confined much more closely to the north polar region than HCN. This confinement is due to the mixing barrier caused by the north polar vortex, which effectively prevents cross latitude mixing and confines species within the vortex core.
Other short lifetime gases (e.g. C$_4$H$_2$) are also observed to be largely confined to the vortex core. HCN has a more complex behaviour, i.e. a gradual south-to-north gradient. This is largely explicable by a combination of HCN's long photochemical lifetime and steep vertical gradient. The steep gradient means it is sensitive to vertical motion - including upwelling in the southern summer hemisphere - whereas its long lifetime means it has time to escape the vortex mixing barrier and be transported to lower latitudes.

The high enrichment of many trace species within the polar vortex provides the possibility of unique chemistry, as in Earth's antarctic vortex, however this has so far been difficult to observe with CIRS due to the (assumed) complexity of any chemical products and cold temperatures.
However, it is possible to use observed polar enrichments to probe the relative lifetimes of many species. If the only processes occurring are known photochemical reactions and vertical atmospheric motion, then the observed north polar enrichment should be proportional to the predicted vertical gradient, i.e. inversely proportional to the predicted species lifetime.

\begin{figure}[t]
\sidecaption
\includegraphics[scale=.35]{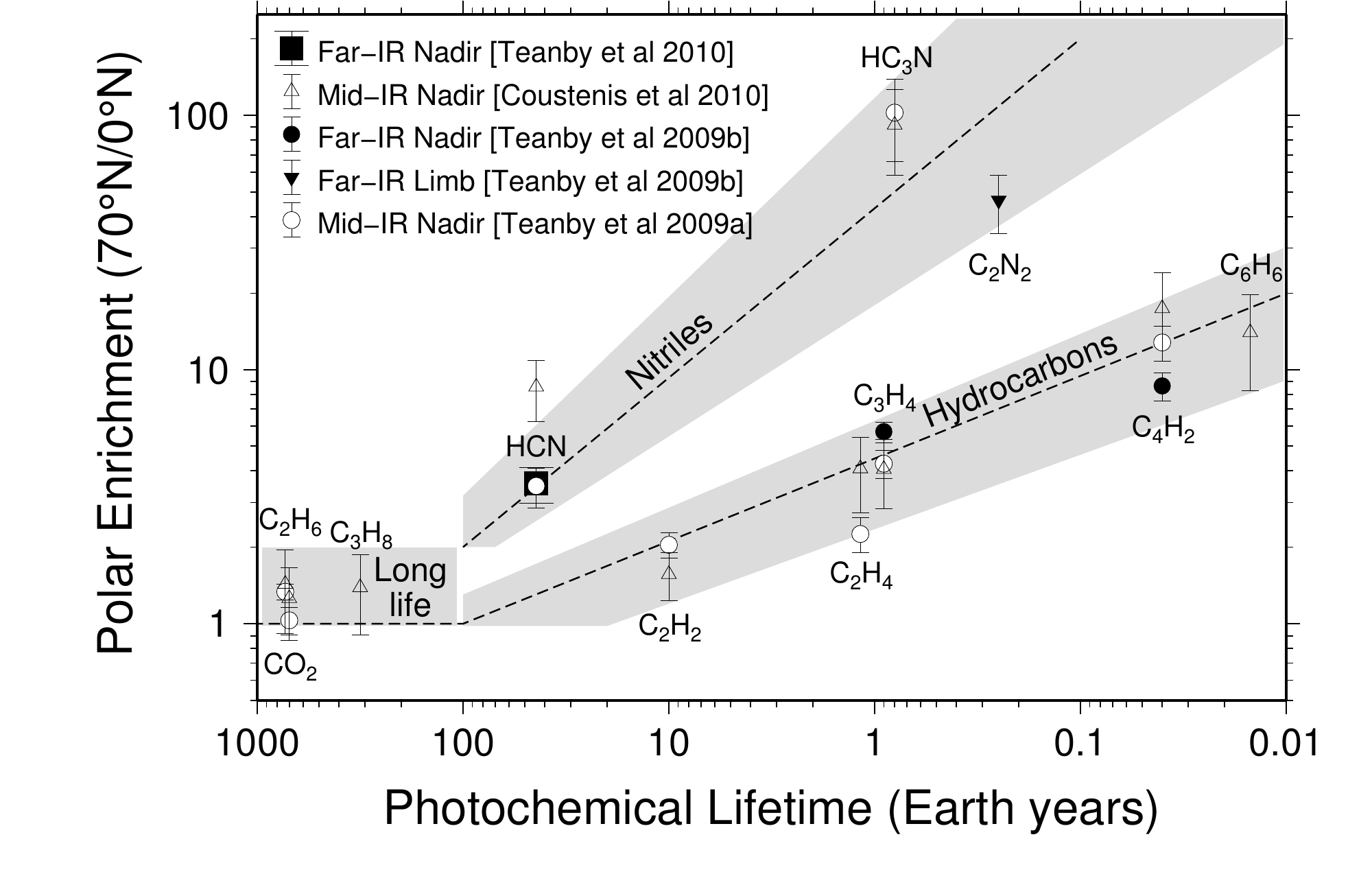}
\caption{Observed polar enrichment as a function of predicted photochemical lifetime from \cite{04wilatr}. Shorter lifetime species are typically more enriched, although nitriles appear to be anomalously enriched compared to hydrocarbons. This suggests an additional loss mechanism for nitrile species. Redrawn from \cite{10teaetal_faraday}.}
\label{fig:rats} 
\end{figure}

Fig.~\ref{fig:rats} shows there is indeed such a relation, but surprisingly the nitrile trend appears separate to that for hydrocarbons.
This suggests that there is some additional sink mechanism causing the vertical gradient of nitriles to be steeper than expected from photochemical models.
The cause of this is currently unknown, but could include incorporation of nitriles into photochemical hazes or polymerisation \cite{10teaetal_faraday}.

\subsection{Nitrile profiles in the northern winter polar vortex}
\label{sect:vortex}

\begin{figure}[t]
\sidecaption
\includegraphics[scale=.5]{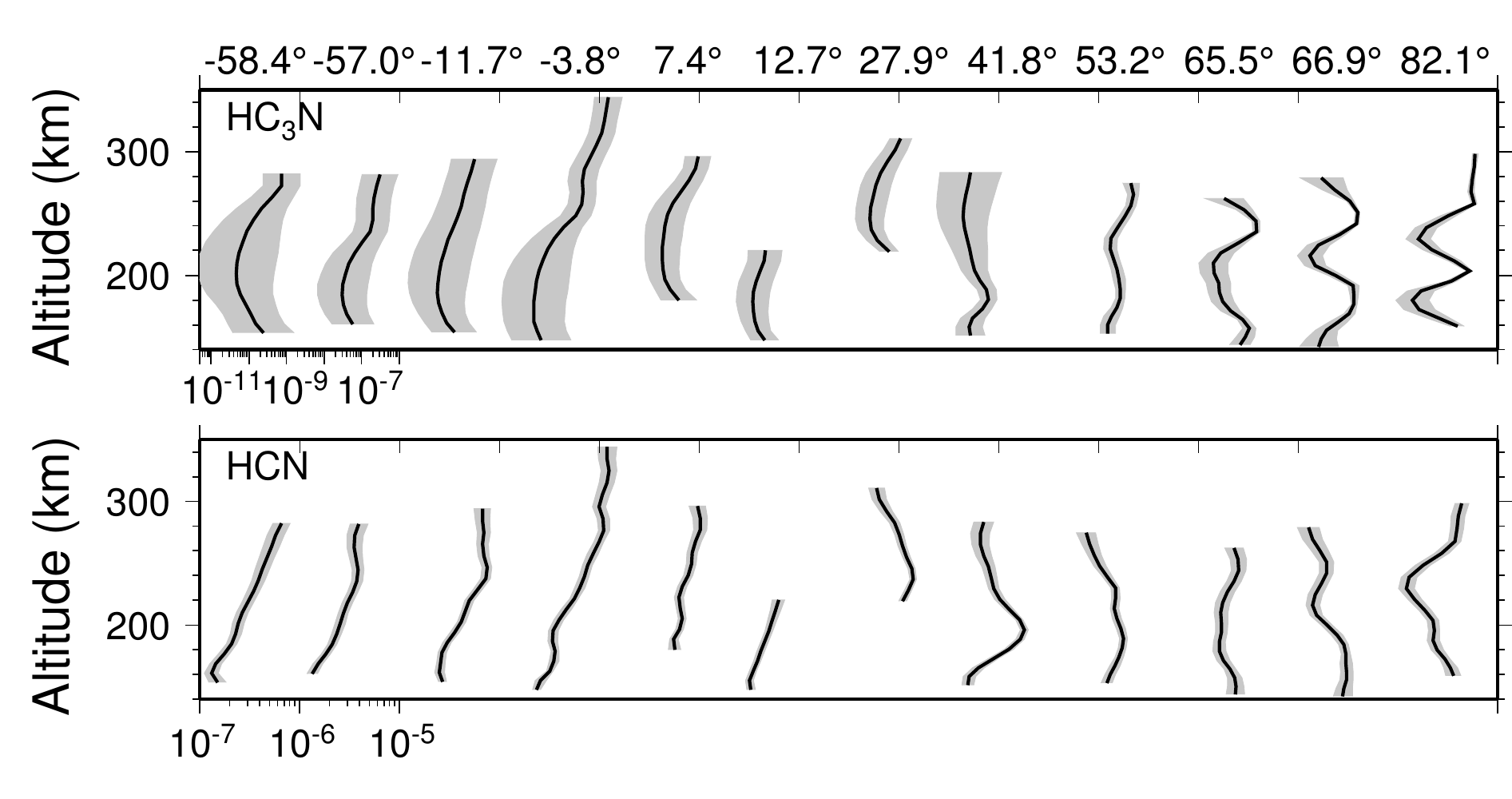}
\caption{Profiles of HCN and HC$_3$N derived from high spatial resolution limb data. Latitudes are given at the top and profiles from different observations are offset horizontally for clarity. North of 60$^{\circ}$N - within the north polar vortex - large amplitude composition layers are evident. These could suggest mixing/instabilities on the vortex boundary allowing mixing of polar and non-polar airmasses. Redrawn from \cite{09teaetal_layers}.}
\label{fig:layers} 
\end{figure}

Specific to the north polar regions are large amplitude composition layers in most trace gases, especially in the nitriles (Fig.~\ref{fig:layers}). These layers are somewhat puzzling and could be linked to discrete haze layers observed by Cassini's Imaging Science Sub-System \cite{05poretal}. Suggested causes of these layers include: chemical sinks, gravity waves, incorporation into haze layers, or dynamics. 

The most plausible of these is a possible dynamical origin. \cite{09teaetal_layers} proposed that the layering was caused by cross vortex mixing, which allows polar and non-polar air masses to mix. A similar process occurs for ozone at the boundary of Earth's polar vortices \cite{95ors}. In this scenario the high wind shear in the vortex causes instabilities and trace-gas poor air is transported across the vortex boundary. Conversely, displaced trace-gas rich air escapes the vortex and is mixed to non-polar atmosphere latitudes. Because the stratosphere is stably stratified, the resulting layers can be fairly long lived and persist long enough to be observed by CIRS limb observations.

If this mechanism is correct then we would expect gases with the most polar enrichment to have the largest amplitude layers, and all the gas layers for the same observation to have similar altitudes, which both appear to be the case (Fig.~\ref{fig:layers}). This would not be the case if a chemical sink were responsible, unless it acted on all gases similarly, which seems unlikely. Also, a gravity wave origin does not seem to provide large enough amplitude to explain up to 50-fold layer amplitudes for HC$_3$N. However, it is thought that gravity waves play a role in eroding the vortex wall and contributing to the cross latitude mixing \cite{08teaetal_jgr}. It can be seen that nitrile chemistry and atmospheric dynamics appear to be inextricably intertwined on Titan. Further insight will be possible as the Cassini mission progresses.

\section{Nitrile condensables in Titan's stratosphere}
\label{sect:ices}

In this section, we discuss the detection of nitrile ices observed in Titan's stratosphere by CIRS. Given that CIRS is the successor to the IRIS instrument onboard Voyager, we will provide a brief overview of the ices observed by IRIS, which motivated the continued search for stratospheric ices with CIRS, nearly one Saturn year ($\sim$ 30 terrestrial years) later.

\subsection{Nitrile ice cloud characteristics in the thermal infrared}
\label{sect:clouds}

There are two distinct types of cloud systems in Titan's atmosphere. The first is condensed \methane, typically found in Titan's troposphere and similar to water clouds in Earth's troposphere. The second type of ice cloud arrises in Titan's much more dynamically stable stratosphere and is composed of condensed nitriles and/or hydrocarbons. The latter type are the clouds that CIRS detects in Titan's stratosphere.

\begin{figure}[b]
\includegraphics[scale=.175]{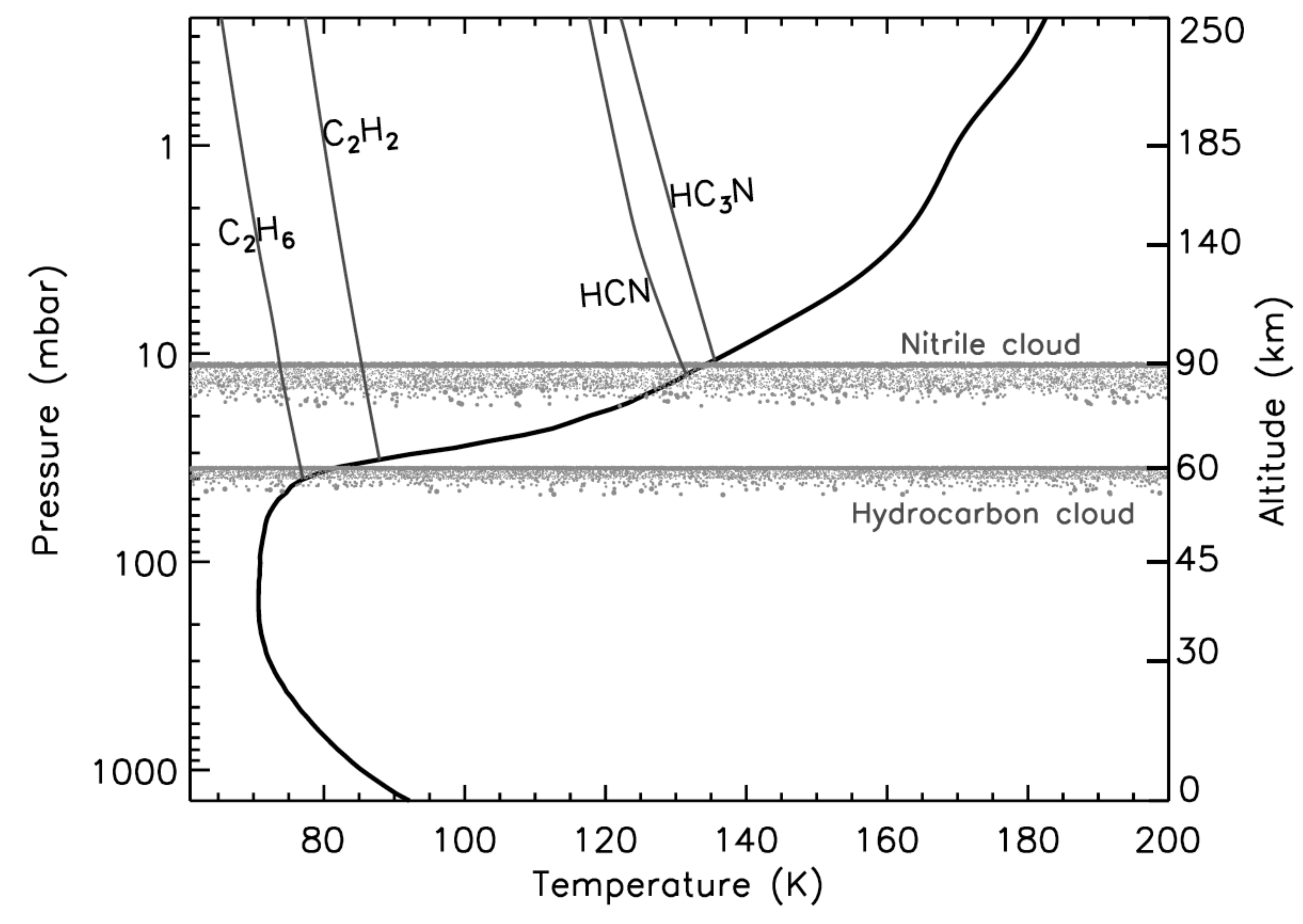}
\caption{Figure from \cite{Westbook12}. Temperature dependence of Titan's pressure-altitude relation (thick black curve) at latitude 15$^{\circ}$S in mid-2006 \cite{08achetal, Anderson11}. Superimposed are the derived saturation vapor pressure curves for the four most abundant trace organics (dark grey curves). Vertical distribution of vapor abundances for hydrocarbons and nitriles were patterned after \cite{10vinetal} and \cite{07teaetal}, respectively. Condensation is expected to occur at altitudes below the intersection of the saturation vapor pressure curves with Titan's temperature structure. The two narrow layers represent the altitude locations of generic nitrile and hydrocarbon cloud regions where saturation can occur.
}
\label{fig:satcurv}
\end{figure}

The fate of most organic vapors in Titan's lower stratosphere is condensation, forming sharply layered upper boundaries near the altitude location where condensation is expected. Figure \ref{fig:satcurv}, originally published in \cite{Westbook12}, illustrates the altitude locations that are expected for such ice clouds, when the temperature structure is at 15$^{\circ}$S during mid northern winter on Titan (circa 2006).

Whereas the IRIS low wavenumber cut-off occurred at 200 \cm, CIRS extends further to 10 \cm, covering a significant portion of the sub-millimeter spectrum that turns out to be extremely important and uniquely qualified for the detection of ice signatures from nitriles. In the far-IR between 70 and 270 \cm, nitrile ices reveal numerous overlapping broad-emission features caused by low-energy lattice vibrations \cite{Moore10}. CIRS is able to detect these signatures due to 1) its excellent signal-to-noise in this part of the far-IR spectrum and 2) the Planck intensity is quite strong in this spectral region. Hydrocarbon ices have very small absorption cross-sections in this part of the far-IR, since they tend to not have strong spectral features in this region. This hinders CIRS from detecting hydrocarbon condensate signals, even though their abundances are larger than those of condensed nitriles \cite{07couetal}.

\subsection{CIRS observations of nitrile ices}
\label{sect:observations}

\begin{figure}[h]
\includegraphics[scale=.15]{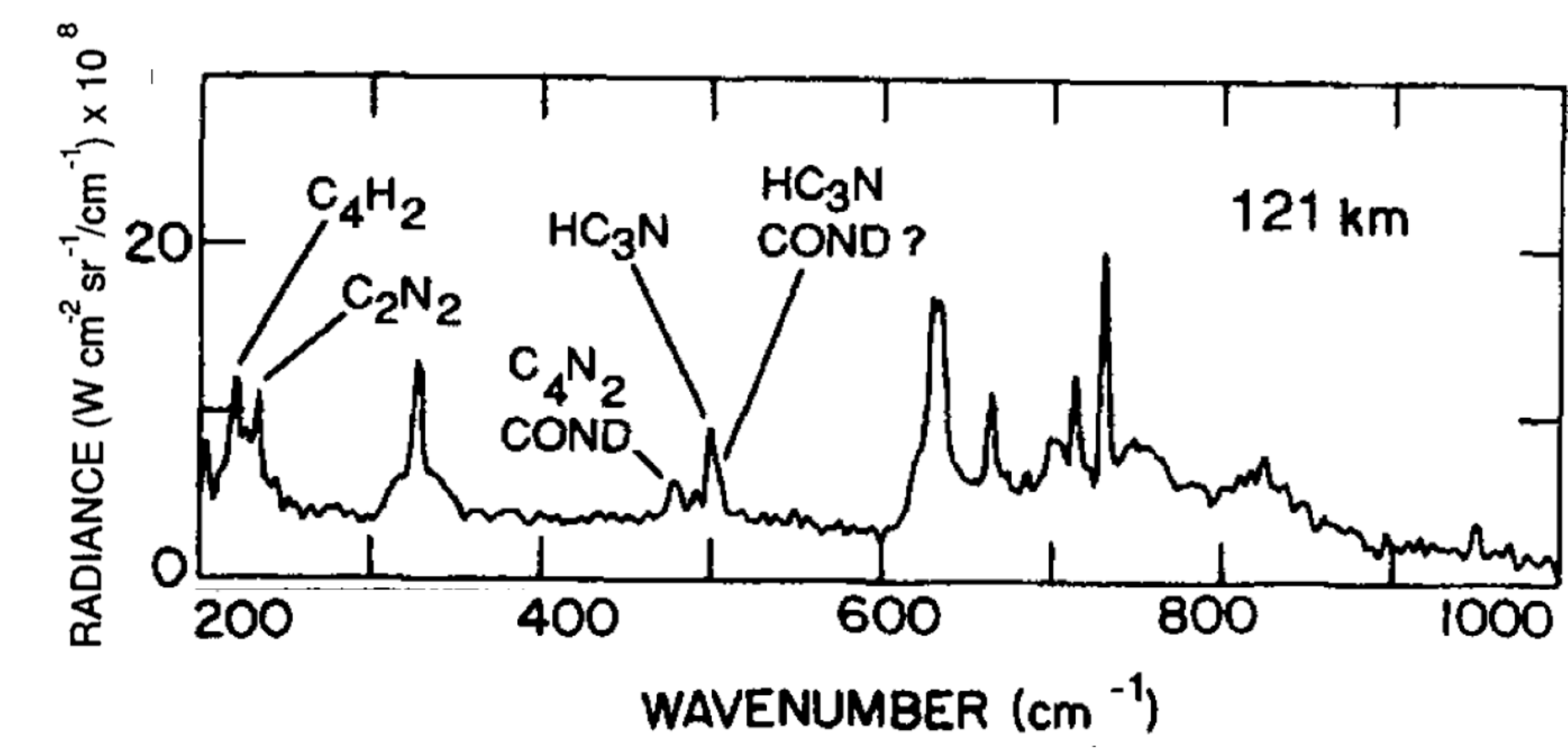}
\caption{Figure adapted from \cite{Samuelson97ice}. Titan's radiance spectrum as observed by IRIS. The spectrum is an average of 3 spectra and was recorded at an altitude of 121 km above Titan's surface horizon (limb-viewing mode). The discovery of the $\nu_{6}$ band of crystalline \cyanoacet $\;$ at 506 \cm $\,$ and the putative $\nu_{8}$ band of crystalline C$_{4}$N$_{2}$ at 478 \cm $\,$ are easily seen.}
\label{fig:iris}
\end{figure}

Sharp ice emission features above Titan's thermal-IR continuum are the easiest to detect provided the abundances are sufficiently large. These features point uniquely to a specific, isolated pure ice (no mixtures). Examples of such features are the $\nu_{6}$ band of crystalline \cyanoacet $\;$ at 506 \cm and the $\nu_{8}$ band of crystalline C$_{4}$N$_{2}$ at 478 \cm, both of which have been observed and identified in Titan's atmosphere from first Voyager IRIS then Cassini CIRS. Figure \ref{fig:iris} is an IRIS spectrum depicting both the \cyanoacet $\;$ condensate at 506 \cm\ and the C$_{4}$N$_{2}$ condensate at 478 \cm\ \cite{Samuelson97ice}. Since the IRIS spectral resolution of 4.3 \cm $\;$ was too low to spectrally separate the \cyanoacet $\;$ condensate at 506 \cm $\;$ from the vapor at 499 \cm, an abundance and particle size determination was not possible. However using the higher spectral discrimination of CIRS, both particle size and abundance were determined from observations of Titan at 70$^{\circ}$N and 62$^{\circ}$N during northern winter \cite{Anderson10}. Regarding the feature at 478 \cm\ observed by both IRIS and CIRS, and tentatively attributed to C$_{4}$N$_{2}$ ice based on spectral location \cite{Masterson90}, there is a caveat that the vapor has never been observed in Titan's atmosphere, which is an absolute requirement for the ice to form.\footnote{An upper limit of 9~ppb for C$_4$N$_2$ gas was determined by \cite{dekok08}.} This is a mystery that is still being addressed today.

\begin{figure}[h]
\includegraphics[scale=.4]{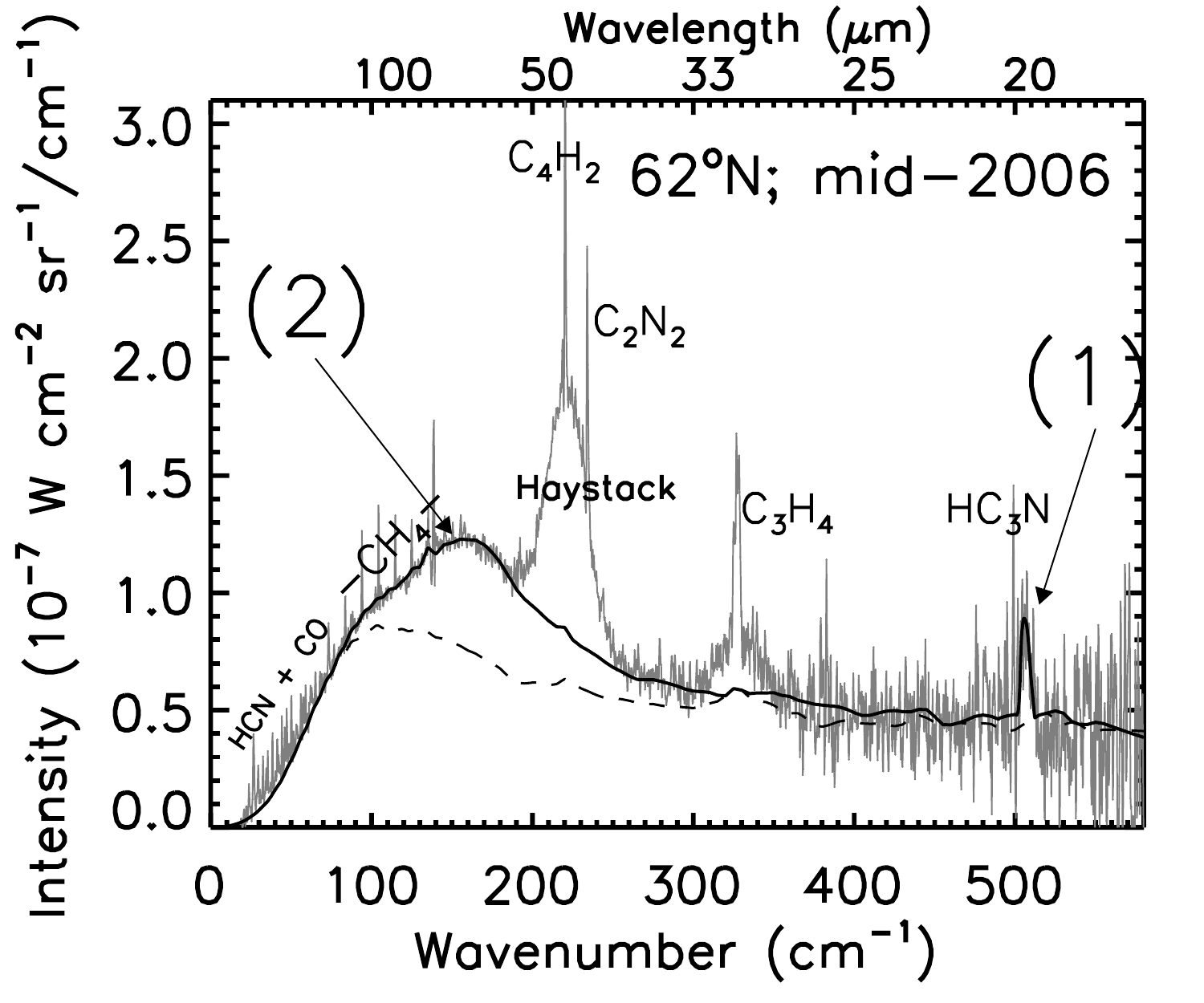}
\caption{Figure and caption from \cite{Westbook12}. CIRS spectrum of Titan at 62$^{\circ}$N (thin black curve). Various organic vapors are labeled as is the unknown solid material has been termed termed the `haystack.' The sharp 506 \cm $\,$ \cyanoacet $\,$ ice emission feature is labeled (1) and the broad 160 \cm $\,$ composite ice feature is labeled (2). The solid and the dashed black curves are respectively synthetic spectra with and without the ice contribution, fit to the data with a radiative transfer analysis. Notice that feature (1) is easily distinguished from the continuum whereas feature (2) is indistinguishable from the continuum and would not be identified as such without a radiative transfer analysis.}
\label{fig:160spec}
\end{figure}

In contrast to the sharp ice emission features that are spectrally easy to detect, ice signatures exist in Titan's far-IR spectrum that are spectrally very broad due to low-energy lattice vibrations (librations), and are comprised of numerous overlapping emissions that are impossible to isolate individually. Even though these types of composite features are intrinsically much stronger than those due to pure ices, they are much more difficult to detect and identify because of their quasi-continuum nature. Thus, the spectral dependence must be derived from a full radiative analysis \cite{Anderson11,dekok07b}. An example of this type of identification is the CIRS-discovered nitrile composite ice feature that peaks at 160 \cm, labeled (2) in Fig. \ref{fig:160spec} (after\cite{Westbook12}.)

\begin{figure}[t]
\sidecaption
\includegraphics[scale=.25]{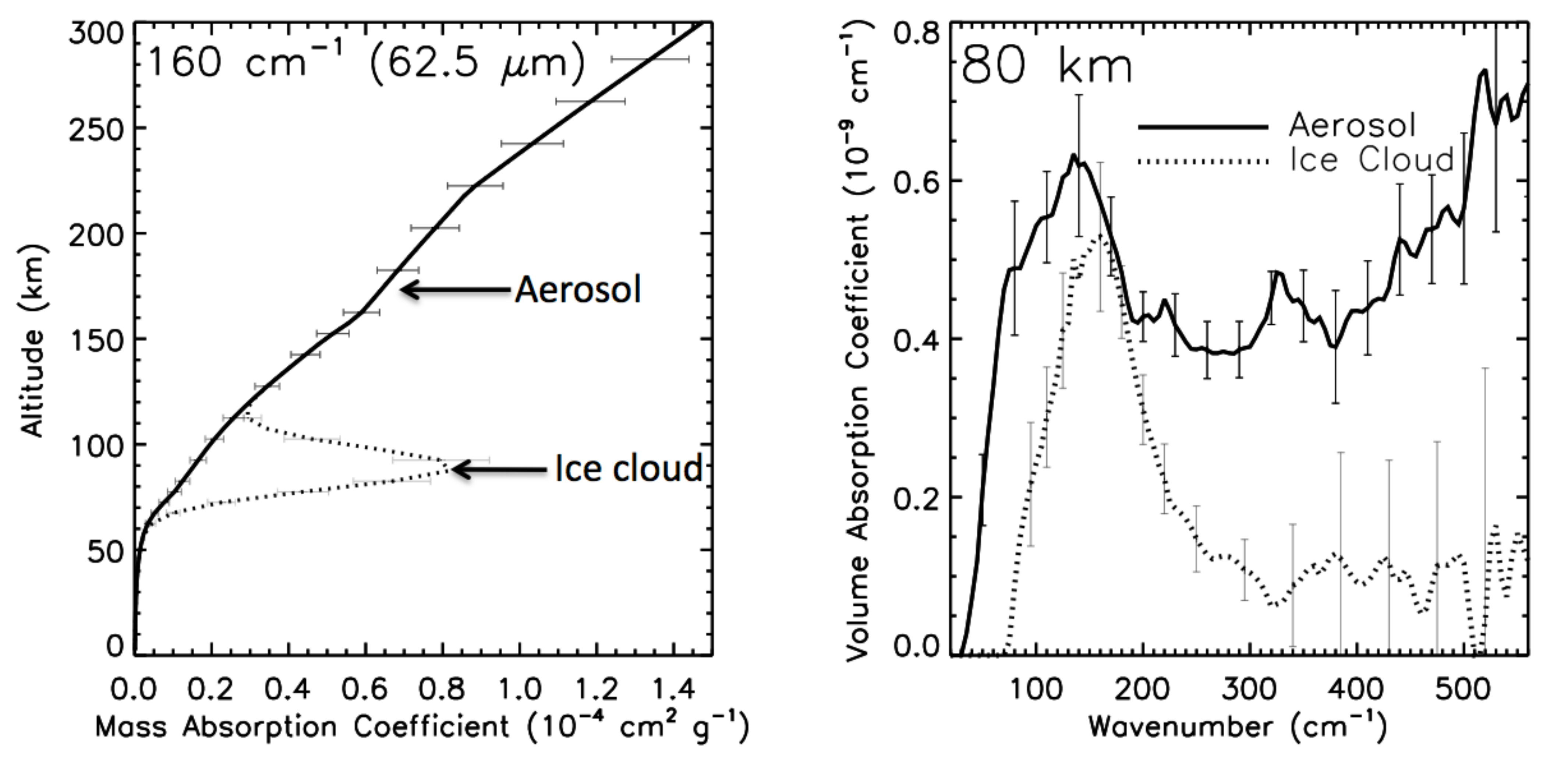}
\caption{Figures and caption adapted from \cite{Anderson11}. Left side: Derived vertical distributions of mass absorption coefficient at 160 \cm $\,$ at 15$^{\circ}$S with 1$\sigma$ uncertainties shown. The model assumes a single aerosol and a single ice cloud component. Absorption coefficients are for aerosol only (solid curve) and aerosol plus ice (dotted curve). Right side: Derived spectral dependence of volume absorption coefficient at the 80 km altitude level at 15$^{\circ}$S with 1$\sigma$ uncertainties shown. Absorption coefficients are for aerosol only (solid curve) and aerosol plus ice (dotted curve).}
\label{fig:15S}
\end{figure}

The CIRS-derived vertical and spectral dependence of this ice feature at 160 \cm $\,$ is illustrated in Fig. \ref{fig:15S} at latitude 15$^{\circ}$S. This ice feature is thought to be comprised of a mix of nitrile ices based on the altitude location of the cloud (see Fig. \ref{fig:satcurv}) and the spectral dependence of the observed ice feature spans wavenumbers where nitriles have numerous overlapping absorption features \cite{Anderson11,Moore10}. 

Our analyzes to date \cite{Anderson11,Westbook12,Anderson10,dekok07b} indicate that during mid to late northern winter on Titan (during the Cassini prime and extended missions) a system of thin nitrile ice clouds extend globally from 85$^{\circ}$N to at least 55$^{\circ}$S. These clouds are tentatively assigned to composites of HCN and \cyanoacet, but most likely contain additional trace nitrile ices, based on the broad emission feature seen by CIRS centered at 160 \cm. These ice clouds are found at altitudes around 90 km at equatorial and southern latitudes, and increase in altitude as Titan's temperature structure cools in the stratosphere, and most organic vapor abundances increase during northern winter. We expect these clouds to maintain a global distribution but the abundances will change as functions of latitude as Titan changes season.

\section{Other N-Bearing Species and Isotopes}
\label{sect:others}

In this section, we discuss the search for, and measurement of, very low mixing ratio nitrogen-bearing species: the \hcnfifteen\ isotopologue of HCN, and the species \ammonia\ and \acetonitrile\ that have not yet been detected by CIRS in Titan's atmosphere, although detected previously by other techniques. 

\subsection{$^{15}$N Isotopologues}
\label{sect:isotopes}

The study of isotopic ratios in Titan's main molecular reservoirs of nitrogen and carbon, namely \nitrogen\ and \methane , bring information on the origin and evolution mechanisms of the atmosphere. In contrast, the study of the isotopic ratios in HCN, which is a product of the photodissociation of \methane\ and \nitrogen\ (see Sect. \ref{sect:chem}), gives us information on the atmospheric chemistry. HC$^{15}$N was first detected from ground base millimeter telescopes \cite{97hidetal, 02maretal, 04gur} from the 88.6 and 258.16 GHz lines. The inferred \nratio\ disk averaged isotopic ratios varied between 60.5 and 94. 

Subsequently CIRS detected for the first time the 711 \cm\ band of HC$^{15}$N \cite{vinatier07b}. Limb observations were used to constrain the \nratio\ isotopic ratio in HCN at 15\dg S and 83\dg N for altitudes between 150 and 400 km. Using a constant-with-height vertical isotopic ratio in model calculations allowed the CIRS observations to be reproduced with values of $56^{+16}_{-13}$ and $56^{+10}_{-9}$ at 15\dg S and 83\dg N respectively. As the isotopic ratio does not vary with latitude, it is possible to derive a mean isotopic ratio of $56\pm 8$. 

This derived isotopic ratio in HCN from CIRS data is substantially lower than that measured for N$_2$ (\nratio\ = $167.7\pm 0.6$) in situ with the Huygens Gas Chromatograph/Mass Spectrometer (GCMS) \cite{niemann10}, implying that  HCN is enriched in $^{15}$N compared to \nitrogen . As \nitrogen\ is the main nitrogen reservoir in TitanÕs atmosphere, its greater \nratio\ ratio compared to HCN suggests the existence of a fractionation process in the formation of HCN (daughter) from \nitrogen\ (parent) preferring $^{15}$N over $^{14}$N.\footnote{Or the reverse: preferential destruction of $^{14}$N-bearing HCN.} A likely explanation has been posited \cite{Liang_2007} involving isotope-selective \nitrogen\ photodissociation. This process occurs at wavelengths shorter than 100 nm, mainly through predissociation transitions toward Rydberg and valence states \cite{Haverd_2005, Lewis_2005, Stark_2005}. These states have long enough lifetimes to display rotational and vibrational structures. The vibrational bands of the two isotopes $^{14}$N$^{15}$N and $^{15}$N$^{15}$N can be shifted by tens of wavenumbers compared to the $^{14}$N$_2$ bands \cite{ Lewis_2005,Sprengers_2003}. As $^{14}$\nitrogen\ is the most abundant isotope, it absorbs most of the solar radiation in the high atmosphere, while the radiation able to photodissociate $^{14}$N$^{15}$N and $^{15}$N$^{15}$N will penetrate at deeper levels. Therefore, $^{14}$N$^{15}$N and $^{15}$N$^{15}$N can be photo-dissociated at deeper levels than most of $^{14}$\nitrogen . This process can therefore greatly increases the \nratio\ isotopic ratio in HCN compared to \nitrogen .

\subsection{The search for further N-bearing molecules with CIRS}
\label{sect:search}

In this subsection we discuss the search for \ammonia\ and \acetonitrile\ in Titan's stratosphere, both of which are expected to be present at low levels from photochemical models. Both ammonia and acetonitrile have been directly detected in Titan's ionosphere by the Cassini mass spectrometer (INMS) \cite{vuitton07}. However, only acetonitrile has been previously detected in the stratospheric/mesospheric region of the atmosphere sensed by CIRS, and using ground-based sub-millimeter telescopes \cite{02maretal} rather than the infrared spectral region.

Therefore, a recent study by the CIRS team attempted to search rigorously for these species in the high S/N limb spectra of Titan acquired specifically for the purpose \cite{nixon10b}. Data from long limb integrations (4 hours) from two flybys (T55 at 25\dg S, and T64 at 76\dg N) were originally analyzed: since then one further observation has been analyzed as described below. The bands used were located in the 9--11~\micron\ spectral region, where Titan's infrared spectrum is relatively uncluttered by the strong alkane and alkyne bands seen elsewhere, and CIRS has high sensitivity. For \ammonia , the $\nu_2$ band centered at 950~\cm\ was employed, while the $\nu_7$ band of \acetonitrile\ at 1041~\cm\ was likewise selected. 

The line list for ammonia was already available in a standard atlas (HITRAN 2008) \cite{rothman09}, while the requisite line list for acetonitrile was created for the study. Using these line lists, synthetic Titan spectra were calculated, incrementally adding quantities of each gas to the model atmosphere, until the gas emission signature exceeded the noise threshold by 1, 2, and 3-$\sigma$ amounts. The results table of \cite{nixon10b} is reproduced here, adding an additional data line for acetonitrile, by analysis of spectra acquired during the more recent T72 flyby (September 24th 2010) that has been computed since the 2010 publication. This adds additional constraint for \acetonitrile\ at 0.30 mbar in the stratosphere at 76~\dg N, a lower altitude than previously.

\begin{table}
\caption{Upper limits on the abundances of \ammonia\ and \acetonitrile\ in Titan's atmosphere}
\label{tab:ulimits}      

\begin{tabular}{p{1.2cm}p{1cm}p{1.7cm}p{1cm}p{1.2cm}p{1cm}p{1cm}p{1cm}}
\hline\noalign{\smallskip}
Gas & Flyby & Date & Lat. & Pressure &  \multicolumn{3}{c}{VMR Upper Limit (ppbv)} \\
Name & No. & & (\dg ) & (mbar) & 1-$\sigma$ & 2-$\sigma$ & 3-$\sigma$ \\
\noalign{\smallskip}\svhline\noalign{\smallskip}
\ammonia\  & T55 & 22-MAY-09 & 25\dg S & 7.6 & 0.59 & 0.88 & 1.3 \\
                     & T64 & 28-DEC-09 & 76\dg N & 0.26 & 2.0 & 6.4 & 14 \\
\noalign{\smallskip}
\acetonitrile\ & T55 & 22-MAY-09 & 25\dg S & 0.27 & 49 & 78 & 109 \\
                       & T64 & 28-DEC-09 & 76\dg N & 0.018 & 660 & 830 & 1000 \\
                       & T72 & 24-SEP-10 & 76\dg N & 0.30 & 53 & 70 & 89 \\
\noalign{\smallskip}\hline\noalign{\smallskip}
\end{tabular}
\end{table}

\section{Conclusions and future directions}
\label{sect:conc}

In this chapter we have described the chemical origin and stratospheric distributions of nitrogen-bearing molecules in Titan's stratosphere as observed by Cassini CIRS, focusing on the infrared-detected nitrile gases HCN, HC$_3$N and C$_2$N$_2$. Measurements of the abundances of these gases is important not only for constraining and improving models of the chemistry, but also because their different lifetimes allow us to track atmospheric motions.

At the beginning of the Cassini era in 2004, all three nitriles exhibited steep positive vertical gradients of abundance at low latitudes. In contrast, at the high northern latitudes then experiencing winter, the profiles were more nearly constant with height implying that strong subsidence was occurring - firm evidence of the large global Hadley circulation cell that had been predicted by dynamical models. The details differed for the three gases, with the longer-lived HCN showing positive enhancement even at low altitudes towards the mid-latitudes: evidence of the returning branch of the circulation cell.

The long duration of the Cassini mission (8 years in Saturn orbit and counting) is now permitting us to observe the turning of Titan's long seasons, as the northern enrichments have begun to exhibit  fading, even as abundance enhancements have begun to emerge in Titan's far south. We may look forward to witnessing further details of this changing circulation during the latter years of the Cassini mission - hoped to last until 2017 near the time of northern summer solstice.

Even as we now gain some confidence in our growing knowledge of this picture, mysteries remain that continue to elude our understanding. In particular, we remark on the curious difference in the trends of enhancement versus lifetime for the nitriles shown in Fig.~\ref{fig:rats} compared to the hydrocarbons - with the implication that there are processes occurring that are not yet included in our models. To this list we could add the current non-detection of CH$_3$CN in the stratosphere by CIRS (but seen by ground-based telescopes), or the similar elusiveness of NH$_3$ - firmly detected in the upper atmosphere. New missions and instruments may be required to finally resolve these and other riddles.


\begin{thebibliography}{10}
\providecommand{\url}[1]{{#1}}
\providecommand{\urlprefix}{URL }
\expandafter\ifx\csname urlstyle\endcsname\relax
  \providecommand{\doi}[1]{DOI \discretionary{}{}{}#1}\else
  \providecommand{\doi}{DOI \discretionary{}{}{}\begingroup
  \urlstyle{rm}\Url}\fi

\bibitem{niemann10}
H.B. {Niemann}, S.K. {Atreya}, J.E. {Demick}, D.~{Gautier}, J.A. {Haberman},
  D.N. {Harpold}, W.T. {Kasprzak}, J.I. {Lunine}, T.C. {Owen}, F.~{Raulin},
  Journal of Geophysical Research (Planets) \textbf{115}, 12006 (2010).
\newblock \doi{10.1029/2010JE003659}

\bibitem{hanel81}
R.~{Hanel}, B.~{Conrath}, F.M. {Flasar}, V.~{Kunde}, W.~{Maguire}, J.C.
  {Pearl}, J.~{Pirraglia}, R.~{Samuelson}, L.~{Herath}, M.~{Allison}, D.P.
  {Cruikshank}, D.~{Gautier}, P.J. {Gierasch}, L.~{Horn}, R.~{Koppany},
  C.~{Ponnamperuma}, Science \textbf{212}, 192 (1981).
\newblock \doi{10.1126/science.212.4491.192}

\bibitem{kunde81}
V.G. {Kunde}, A.C. {Aikin}, R.A. {Hanel}, D.E. {Jennings}, W.C. {Maguire}, R.E.
  {Samuelson}, {Nature} \textbf{292}, 686 (1981).
\newblock \doi{10.1038/292686a0}

\bibitem{maguire81}
W.C. {Maguire}, R.A. {Hanel}, D.E. {Jennings}, V.G. {Kunde}, R.E. {Samuelson},
  Nature \textbf{292}, 683 (1981).
\newblock \doi{10.1038/292683a0}

\bibitem{nixon10b}
C.A. {Nixon}, R.K. {Achterberg}, N.A. {Teanby}, P.G.J. {Irwin}, J.M. {Flaud},
  I.~{Kleiner}, A.~{Dehayem-Kamadjeu}, L.R. {Brown}, R.L. {Sams},
  B.~{B{\'e}zard}, A.~{Coustenis}, T.M. {Ansty}, A.~{Mamoutkine},
  S.~{Vinatier}, G.L. {Bjoraker}, D.E. {Jennings}, P.N. {Romani}, F.M.
  {Flasar}, Faraday Discussions \textbf{147}, 65 (2010).
\newblock \doi{10.1039/c003771k}

\bibitem{waite04}
J.H. {Waite}, W.S. {Lewis}, W.T. {Kasprzak}, V.G. {Anicich}, B.P. {Block}, T.E.
  {Cravens}, G.G. {Fletcher}, W.H. {Ip}, J.G. {Luhmann}, R.L. {McNutt}, H.B.
  {Niemann}, J.K. {Parejko}, J.E. {Richards}, R.L. {Thorpe}, E.M. {Walter},
  R.V. {Yelle}, {Space Sci. Rev.} \textbf{114}, 113 (2004).
\newblock \doi{10.1007/s11214-004-1408-2}

\bibitem{vuitton07}
V.~{Vuitton}, R.V. {Yelle}, M.J. {McEwan}, {Icarus} \textbf{191}, 722 (2007).
\newblock \doi{10.1016/j.icarus.2007.06.023}

\bibitem{flasar04b}
F.M. {Flasar}, V.G. {Kunde}, M.M. {Abbas}, R.K. {Achterberg}, P.~{Ade},
  A.~{Barucci}, B.~{B{\'e}zard}, G.L. {Bjoraker}, J.C. {Brasunas},
  S.~{Calcutt}, R.~{Carlson}, C.J. {C{\'e}sarsky}, B.J. {Conrath},
  A.~{Coradini}, R.~{Courtin}, A.~{Coustenis}, S.~{Edberg}, S.~{Edgington},
  C.~{Ferrari}, T.~{Fouchet}, D.~{Gautier}, P.J. {Gierasch}, K.~{Grossman},
  P.~{Irwin}, D.E. {Jennings}, E.~{Lellouch}, A.A. {Mamoutkine}, A.~{Marten},
  J.P. {Meyer}, C.A. {Nixon}, G.S. {Orton}, T.C. {Owen}, J.C. {Pearl},
  R.~{Prang{\'e}}, F.~{Raulin}, P.L. {Read}, P.N. {Romani}, R.E. {Samuelson},
  M.E. {Segura}, M.R. {Showalter}, A.A. {Simon-Miller}, M.D. {Smith}, J.R.
  {Spencer}, L.J. {Spilker}, F.W. {Taylor}, {Space Sci. Rev.} \textbf{115}, 169
  (2004).
\newblock \doi{10.1007/s11214-004-1454-9}

\bibitem{lavvas08a}
P.~Lavvas, A.~Coustenis, I.~Vardavas, Planetary and Space Science
  \textbf{56}(1), 27  (2008).
\newblock \doi{DOI: 10.1016/j.pss.2007.05.026}.
\newblock
  \urlprefix\url{http://www.sciencedirect.com/science/article/B6V6T-4PN05S4-2/2/2f028ed3f317ff898f73c50d4b73fb21}.
\newblock Surfaces and Atmospheres of the Outer Planets, their Satellites and
  Ring Systems: Part III, European Geosciences Union General Assembly -
  Sessions PS3.02 and PS3.03

\bibitem{kunde96}
V.G. {Kunde}, P.A. {Ade}, R.D. {Barney}, D.~{Bergman}, J.F. {Bonnal},
  R.~{Borelli}, D.~{Boyd}, J.C. {Brasunas}, G.~{Brown}, S.B. {Calcutt},
  F.~{Carroll}, R.~{Courtin}, J.~{Cretolle}, J.A. {Crooke}, M.A. {Davis},
  S.~{Edberg}, R.~{Fettig}, M.~{Flasar}, D.A. {Glenar}, S.~{Graham}, J.G.
  {Hagopian}, C.F. {Hakun}, P.A. {Hayes}, L.~{Herath}, L.~{Horn}, D.E.
  {Jennings}, G.~{Karpati}, C.~{Kellebenz}, B.~{Lakew}, J.~{Lindsay},
  J.~{Lohr}, J.J. {Lyons}, R.J. {Martineau}, A.J. {Martino}, M.~{Matsumura},
  J.~{McCloskey}, T.~{Melak}, G.~{Michel}, A.~{Morell}, C.~{Mosier}, L.~{Pack},
  M.~{Plants}, D.~{Robinson}, L.~{Rodriguez}, P.~{Romani}, W.J. {Schaefer},
  S.~{Schmidt}, C.~{Trujillo}, T.~{Vellacott}, K.~{Wagner}, D.~{Yun}, in
  \emph{Society of Photo-Optical Instrumentation Engineers (SPIE) Conference
  Series}, \emph{Society of Photo-Optical Instrumentation Engineers (SPIE)
  Conference Series}, vol. 2803, ed. by {L.~Horn} (1996), \emph{Society of
  Photo-Optical Instrumentation Engineers (SPIE) Conference Series}, vol. 2803,
  pp. 162--177

\bibitem{brasunas04}
J.C. {Brasunas}, B.~{Lakew}, Recent Res. Devel. Optics \textbf{4}, 95 (2004)

\bibitem{nixon09a}
C.A. {Nixon}, N.A. {Teanby}, S.B. {Calcutt}, S.~{Aslam}, D.E. {Jennings}, V.G.
  {Kunde}, F.M. {Flasar}, P.G. {Irwin}, F.W. {Taylor}, D.A. {Glenar}, M.D.
  {Smith}, {Applied Optics} \textbf{48}, 1912 (2009).
\newblock \doi{10.1364/AO.48.001912}

\bibitem{nixon10c}
C.A. {Nixon}, R.K. {Achterberg}, F.M. {Flasar}, IEEEAC p. 1174 (2010)

\bibitem{nixon12a}
C.A. {Nixon}, T.M. {Ansty}, F.M. {Flasar}, R.K. {Achterberg}, IEEEAC p. 1633
  (2012)

\bibitem{Anderson11}
C.M. {Anderson}, R.E. {Samuelson}, Icarus \textbf{212}, 762 (2011)

\bibitem{07teaetal}
N.A. Teanby, P.G.J. Irwin, R.~de~Kok, S.~Vinatier, B.~B\'{e}zard, C.A. Nixon,
  F.M. Flasar, S.B. Calcutt, N.E. Bowles, L.~Fletcher, C.~Howett, F.W. Taylor,
  Icarus \textbf{186}, 364 (2007)

\bibitem{07vinetal}
S.~Vinatier, B.~B\'{e}zard, T.~Fouchet, N.A. Teanby, R.~de~Kok, P.G.J. Irwin,
  B.J. Conrath, C.A. Nixon, P.N. Romani, F.M. Flasar, A.~Coustenis, Icarus
  \textbf{188}, 120 (2007)

\bibitem{97hidetal}
T.~Hidayat, A.~Marten, B.~B\'{e}zard, D.~Gautier, T.~Owen, H.~Matthews,
  G.~Paubert, Icarus \textbf{126}(1), 170 (1997)

\bibitem{02maretal}
A.~Marten, T.~Hidayat, Y.~Biraud, R.~Moreno, Icarus \textbf{158}, 532 (2002)

\bibitem{04gur}
M.A. Gurwell, Astrophys. J. \textbf{616}, L7 (2004)

\bibitem{09teaetal_c2n2}
N.A. Teanby, P.G.J. {Irwin}, R.~{de Kok}, A.~{Jolly}, B.~B\'{e}zard, C.A.
  {Nixon}, S.B. {Calcutt}, Icarus \textbf{202}, 620 (2009)

\bibitem{04wilatr}
E.H. Wilson, S.K. Atreya, J. Geophys. Res. \textbf{109}, E06002 (2004)

\bibitem{01lebetal}
S.~Lebonnois, D.~Toublanc, F.~Hourdin, P.~Rannou, Icarus \textbf{152}, 384
  (2001)

\bibitem{09teaetal_rs}
N.A. Teanby, P.G.J. {Irwin}, R.~{de Kok}, C.A. {Nixon}, Phil. Trans. R. Soc.
  Lond. A \textbf{367}, 697 (2009)

\bibitem{05flaetal}
F.M. Flasar, R.K. Achterberg, B.J. Conrath, P.J. Gierasch, V.G. Kunde, C.A.
  Nixon, G.L. Bjoraker, D.E. Jennings, P.N. Romani, A.A. Simon-Miller,
  B.~B\'{e}zard, A.~Coustenis, P.G.J. Irwin, N.A. Teanby, J.~Brasunas, J.C.
  Pearl, M.E. Segura, R.C. Carlson, A.~Mamoutkine, P.J. Schinder, A.~Barucci,
  R.~Courtin, T.~Fouchet, D.~Gautier, E.~Lellouch, A.~Marten, R.~Prange,
  S.~Vinatier, D.F. Strobel, S.B. Calcutt, P.L. Read, F.W. Taylor, N.~Bowles,
  R.E. Samuelson, G.S. Orton, L.J. Spilker, T.C. Owen, J.R. Spencer, M.R.
  Showalter, C.~Ferrari, M.M. Abbas, F.~Raulin, S.~Edgington, P.~Ade, E.H.
  Wishnow, Science \textbf{308}, 975 (2005)

\bibitem{06teaetal}
N.A. Teanby, P.G.J. Irwin, R.~de~Kok, C.A. Nixon, A.~Coustenis, B.~B\'{e}zard,
  S.B. Calcutt, N.E. Bowles, F.M. Flasar, L.~Fletcher, C.~Howett, F.W. Taylor,
  Icarus \textbf{181}, 243 (2006)

\bibitem{07couetal}
A.~Coustenis, R.K. Achterberg, B.J. Conrath, D.E. Jennings, A.~Marten,
  D.~Gautier, C.A. Nixon, F.M. Flasar, N.A. Teanby, B.~B\'{e}zard, R.E.
  Samuelson, R.C. Carlson, E.~Lellouch, G.L. Bjoraker, P.N. Romani, F.W.
  Taylor, P.G. Irwin, T.~Fouchet, A.~Hubert, G.S. Orton, V.G. Kunde,
  S.~Vinatier, J.~Mondellini, M.M. Abbas, R.~Courtin, Icarus \textbf{189}, 35
  (2007)

\bibitem{08teaetal}
N.A. Teanby, P.G.J. {Irwin}, R.~{de Kok}, C.A. {Nixon}, A.~{Coustenis},
  E.~{Royer}, S.B. {Calcutt}, N.E. {Bowles}, L.~{Fletcher}, C.~{Howett}, F.W.
  {Taylor}, Icarus \textbf{193}, 595 (2008)

\bibitem{08teaetal_jgr}
N.A. Teanby, R.~de~Kok, P.G.J. Irwin, S.~Osprey, S.~Vinatier, P.J. Gierasch,
  P.L. Read, F.M. Flasar, B.J. Conrath, R.K. Achterberg, B.~Bezard, C.A. Nixon,
  S.B. Calcutt, J. Geophys. Res. \textbf{113}, E12003 (2008)

\bibitem{10couetal}
A.~Coustenis, C.~Nixon, R.~Achterberg, P.~Lavvas, S.~Vinatier, N.~Teanby,
  G.~Bjoraker, R.~Carlson, L.~Piani, G.~Bampasidis, F.~Flasar, P.~Romani,
  Icarus \textbf{207}, 461 (2010)

\bibitem{10teaetal_faraday}
N.A. Teanby, P.G.J. Irwin, R.~de~Kok, C.A. Nixon, Faraday Discussions
  \textbf{147}, 51 (2010)

\bibitem{10teaetal_apj}
N.A. Teanby, P.G.J. Irwin, R.~de~Kok, C.A. Nixon, Astrophys. J. \textbf{724},
  L84 (2010)

\bibitem{10vinetal}
S.~Vinatier, B.~B\'{e}zard, C.A. Nixon, A.~Mamoutkine, R.C. Carlson, D.E.
  Jennings, E.A. Guandique, N.A. Teanby, G.L. Bjoraker, F.M. Flasar, V.G.
  Kunde, Icarus \textbf{205}, 559 (2010)

\bibitem{08achetal}
R.K. Achterberg, B.J. Conrath, P.J. Gierasch, F.M. Flasar, C.A. Nixon, Icarus
  \textbf{194}, 263 (2008)

\bibitem{05poretal}
C.C. Porco, E.~Baker, J.~Barbara, K.~Beurle, A.~Brahic, J.A. Burns, S.~Charnoz,
  N.~Cooper, D.D. Dawson, A.D. Del~Genio, T.~Denk, L.~Dones, U.~Dyudina, M.W.
  Evans, S.~Fussner, B.~Giese, K.~Grazier, P.~Helfenstein, A.P. Ingersoll, R.A.
  Jacobson, T.V. Johnson, A.~McEwen, C.D. Murray, G.~Neukum, W.M. Owen,
  J.~Perry, T.~Roatsch, J.~Spitale, S.~Squyres, P.~Thomas, M.~Tiscareno, E.P.
  Turtle, A.R. Vasavada, J.~Veverka, R.~Wagner, R.~West, Nature \textbf{434},
  159 (2005)

\bibitem{09teaetal_layers}
N.A. Teanby, P.G.J. Irwin, R.~de~Kok, Icarus \textbf{204}, 645 (2009)

\bibitem{95ors}
Y.J. Orsolini, Quart. J. Roy. Met. Soc. \textbf{121}, 1923 (1995)

\bibitem{Westbook12}
R.~{West}, P.~{Lavvas}, C.~{Anderson}, H.~{Imanaka}, in \emph{Titan: Surface,
  Atmosphere and Magnetosphere.}, ed. by {Mueller-Wodarg, I., Griffith, E.,
  Lellouch, e. \& Cravens, T.} ({Cambridge University Press/Cambridge Planetary
  Science Series}, in press)

\bibitem{Moore10}
M.H. {Moore}, R.F. {Ferrante}, W.J. {Moore}, R.~{Hudson}, Astrophys. J. Supp.
  \textbf{191}, 96 (2010)

\bibitem{Samuelson97ice}
R.E. {Samuelson}, L.A. {Mayo}, M.A. {Knuckles}, R.J. {Khanna}, Plan. \& Space
  Sci. \textbf{45}, 941 (1997)

\bibitem{Anderson10}
C.M. {Anderson}, R.E. {Samuelson}, G.L. {Bjoraker}, R.K. {Achterberg}, Icarus
  \textbf{207}, 914 (2010)

\bibitem{Masterson90}
C.M. {Masterson}, R.K. {Khanna}, Icarus \textbf{83}, 83 (1990)

\bibitem{dekok08}
R.~{de Kok}, P.G.J. {Irwin}, N.A. {Teanby}, {Icarus} \textbf{197}, 572 (2008)

\bibitem{dekok07b}
R.~{de Kok}, P.G.J. {Irwin}, N.A. {Teanby}, C.A. {Nixon}, D.E. {Jennings},
  L.~{Fletcher}, C.~{Howett}, S.B. {Calcutt}, N.E. {Bowles}, F.M. {Flasar},
  F.W. {Taylor}, {Icarus} \textbf{191}, 223 (2007).
\newblock \doi{10.1016/j.icarus.2007.04.003}

\bibitem{vinatier07b}
S.~{Vinatier}, B.~{B{\'e}zard}, C.A. {Nixon}, Icarus \textbf{191}, 712 (2007).
\newblock \doi{10.1016/j.icarus.2007.06.001}

\bibitem{Liang_2007}
M.C. {Liang}, A.N. {Heays}, B.R. {Lewis}, S.T. {Gibson}, Y.L. {Yung},
  {Astrophys. J. Lett.} \textbf{664}, L115 (2007).
\newblock \doi{10.1086/520881}

\bibitem{Haverd_2005}
V.E. {Haverd}, B.R. {Lewis}, S.T. {Gibson}, G.~{Stark}, {J. Chem. Phys.}
  \textbf{123}, 214304 (2005).
\newblock \doi{10.1063/1.2134704}

\bibitem{Lewis_2005}
B.R. {Lewis}, S.T. {Gibson}, W.~{Zhang}, H.~{Lefebvre-Brion}, J.M. {Robbe}, {J.
  Chem. Phys.} \textbf{122}, 144302 (2005).
\newblock \doi{10.1063/1.1869986}

\bibitem{Stark_2005}
G.~{Stark}, K.P. {Huber}, K.~{Yoshino}, P.L. {Smith}, K.~{Ito}, {J. Chem.
  Phys.} \textbf{123}, 214303 (2005).
\newblock \doi{10.1063/1.2134703}

\bibitem{Sprengers_2003}
J.P. {Sprengers}, W.~{Ubachs}, K.G.H. {Baldwin}, B.R. {Lewis}, W.{\"U}.L.
  {Tchang-Brillet}, J. Chem. Phys. \textbf{119}, 3160 (2003).
\newblock \doi{10.1063/1.1589478}

\bibitem{rothman09}
L.S. {Rothman}, I.E. {Gordon}, A.~{Barbe}, D.C. {Benner}, P.F. {Bernath},
  M.~{Birk}, V.~{Boudon}, L.R. {Brown}, A.~{Campargue}, J.P. {Champion},
  K.~{Chance}, L.H. {Coudert}, V.~{Dana}, V.M. {Devi}, S.~{Fally}, J.M.
  {Flaud}, R.R. {Gamache}, A.~{Goldman}, D.~{Jacquemart}, I.~{Kleiner},
  N.~{Lacome}, W.J. {Lafferty}, J.Y. {Mandin}, S.T. {Massie}, S.N.
  {Mikhailenko}, C.E. {Miller}, N.~{Moazzen-Ahmadi}, O.V. {Naumenko}, A.V.
  {Nikitin}, J.~{Orphal}, V.I. {Perevalov}, A.~{Perrin}, A.~{Predoi-Cross},
  C.P. {Rinsland}, M.~{Rotger}, M.~{{\v S}ime{\v c}kov{\'a}}, M.A.H. {Smith},
  K.~{Sung}, S.A. {Tashkun}, J.~{Tennyson}, R.A. {Toth}, A.C. {Vandaele},
  J.~{Vander Auwera}, {J. Quant. Spectr. Rad. Trans.} \textbf{110}, 533 (2009).
\newblock \doi{10.1016/j.jqsrt.2009.02.013}

\end{thebibliography}

\end{document}